\documentclass{article}

\begin{document}

\title{Electric-magnetic duality beyond four dimensions and in general 
Relativity 
\footnote{\uppercase{T}his work is supported by \uppercase{CNRS}}}

\author{Bernard L. JULIA \footnote{\uppercase{W}ork done partly in collaboration with 
\uppercase {Y D}olivet, \uppercase{P H L}abord\`ere and \uppercase{L P}aulot resp. with
\uppercase{J L}evie and \uppercase{S R}ay. Based on a talk presented at the 
Chern conference,  Nankai, DGMTP August 2005.}\\
\it Laboratoire de Physique th\'eorique de l'ENS\\
\it 24 rue Lhomond \\ 
\it 75005 Paris, FRANCE\\ 
E-mail: bjulia AT lpt.ens.fr}

\maketitle
\hfill LPT-ENS/05-43
\abstract{After reviewing briefly the classical examples of duality in 
four dimensional field theory 
we present a generalisation to arbitrary dimensions and to p-form
fields. Then we  explain how U-duality may become part of a larger non abelian 
V-symmetry in superstring/supergravity theories. And finally we discuss two 
new results for 4d gravity theory
with a cosmological constant: a new exact gravitational instanton equation and
a surprizing linearized classical duality around de Sitter space.}

\section{Electric-magnetic  duality and self-duality}

\subsection{Gauge fields}
The discrete ($Z_4$) and continuous ($SO(2,R)$) invariances of 
the Maxwell equation and of the gauge fixed Maxwell action \cite{TD} are
a remarkable feature of 4 dimensional electromagnetism in vacuum. The
inclusion of matter requires non trivial topology (like a possibly 
nontrivial U(1) principal bundle)
in order to preserve these symmetries. At the quantum level the lattice of 
electric-magnetic charges breaks the symmetry down to a discrete one.
The Dirac-Schwinger quantization condition constrains the possible  
charges of a pair of dyons D(e,g) and D'(e',g') to satisfy:
\begin{equation}
4\pi(eg'-e'g)/h \, = \,{\it  integer}\label{DS}
\end{equation}
The two helicities of the electromagnetic field correspond to self-dual and 
antiself-dual field strengths.
In euclidean signature the (real) classical field strength can be decomposed 
locally into the sum of a self-dual part and an anti self-dual part ie 
\begin{equation}
dA\equiv F \, , \, F_\pm=\pm *F_\pm\label{IM}
\end{equation}
where * is the Hodge dualization operator on two forms.
For Yang-Mills fields there is a celebrated generalization of the self-duality 
projection namely the instanton equation. Note that the usual
instanton equation is first order 
and provides only special solutions to the full (vacuum) Yang-Mills equations.

\subsection{Gauge forms}
Pointlike electric charges are minimally coupled to ``vector'' potentials and the 
generalization for scalar fields resp higher (p+1)-form potentials is their coupling 
respectively to instantons and p-branes. Abelian self-duality is possible in even 
dimensional spacetimes of dimension (2p+4) of the appropriate signature for a single 
(p+1)-form potential. There is a generalization of the quantization condition 
(\ref{DS}) to this situation as well and interestingly it involves a plus sign 
rather than a minus sign in (4k+2) dimensions \cite{HD}.

One key property of these remarkable self-dual solutions is that they minimize the action
by saturating a topological charge bound: the so-called BPS bound. It is E. Bogomolny 
who analyzed systematically this mechanism and applied it to magnetic monopoles and 
dyons (independently studied by M. Prasad and C. Sommerfield) .
The lower bound is typically a characteristic (for instance Pontryagin) number of the 
principal bundle under study \cite{EGH}.

\section{U-duality: selecta}

\subsection{Gravity case}
The Einstein action in D dimensions is invariant under diffeomorphisms of the manifold
$M_D$. Upon dimensional reduction by r commuting one parameter isometry groups the 
effective action on the (D-r) quotient space (of orbits) the set of equations becomes 
invariant under a group of internal symmetries that grows with r. Part of it is expected 
for instance $GL(r,R)$ or at least $SL(r,R)$ but other parts of it come as surprises, the 
first of which is the so-called Ehlers symmetry $SL(2,R)$ that is easy to verify after 
reducing ordinary Einstein 
gravity in $D=4$ by one dimension (r=1). More generally reduction of pure gravity from D 
to 3 dimensions leads to a generalised Ehlers symmetry $SL(D-2,R)$, see for instance 
\cite{CJLP3}. This is a major mystery and constitutes one of our motivations to concentrate 
on dualities in general, to discover new ones and to study their properties.  

\subsection{The supergravity magic triangles}
If one considers at first the internal symmetries (commuting with the Poincar\'e group) 
one encounters often  coset spaces, even  Riemannian symmetric spaces, on which 
these symmetries act as real Lie groups. 
These cosets are the target spaces where scalar fields (ie 0-forms) take 
their values. The symmetries are called U-dualities for historical reasons \cite{HuT}, 
approximately half of them act by (Hodge) dualities on the p-forms in their self-duality 
dimension. A remarkable collection of (pure in D=4) supergravity theories as well as their 
dimensional reductions down to 3 dimensions and their higher dimensional ancestors fit 
into a triangle with partial symmetry under the exchange of the space-time dimension with 
the number of supercharges see \cite{CJLP3}. 
These groups are expected to play an important role in string theory after being broken 
down to a discrete  (arithmetic) subgroup. 

In the  example of 4 dimensions for instance the 
U-duality group of maximal supergravity is the split real form of $E_7$ it contains a parity 
conserving subgroup $SL(8,R)$  and the other generators are dualities. The maximal compact 
subgroup of this real form of $E_7$ is $SU(8)$ sometimes called R-symmetry just to confuse us.
The string ``gauge group'' is expected to be the intersection of the split $E_7$ with the 
discrete group $Sp(56,Z)$. $E_7$ is indeed a subgroup of $Sp(56,R)$.
One must double the number of vector potentials from 28 to 56 to realize locally the action 
of dualities, it turns out that the doubled set of fields obeys first order equations that are 
now equivalent to the second order original equations. We shall recognize this phenomenon as 
rather general and this will lead us to V-dualities. The doubled set of fist order equations is 
nothing but a (twisted) self-duality condition. For an early discussion of doubling see for 
instance \cite{Z}.
\begin{equation}
E.F=S*E.F\label{SD}
\end{equation}
In our example $F$ is the 56-plet of field strengths, $E$ is a representative of the scalar 
fields taking their values in the exceptional group $E_7$ and written in the {\bf 56} 
representation and $S$ is a pseudo involution of square $\pm 1$ that compensates for the 
square of the Hodge operation $**=\pm 1$. 

\section{V-duality}
\subsection{del Pezzo surfaces and Borcherds algebras}
Another mystery of duality is the occurrence not only of the exceptional group $E_7$ 
but of the full (extended in fact) $E$ series: $E_8$, $E_7$, $E_6$, $E_5=D_5$, $E_4=A_4$, 
$E_3=A_2\times A_1$... both as the U-duality groups of  maximal supergravity reduced to 3,4,5,6,7,8... 
dimensions and as symmetry groups of type II string theories after torus compactifications.
The equally mysterious occurrence of the $E$ groups or rather of
their Weyl groups acting on the middle cohomology of the socalled del Pezzo 
complex surfaces may be a related phenomenon.  There are in
fact two candidates for $E_1$ so let us choose $A_1$ which is known to be associated to 
the trivial bundle $CP^1\times CP^1$ (one of the two ``minimal del Pezzo surfaces''). 
$SL(2,Z)=A_1$ is known to be also the U-duality group of type IIB 
superstring theory in 10 dimensions (the top dimension). Besides the information provided 
by algebraic geometers (Y. Manin...) we used \cite{JHLP1} one important remark of C. Vafa 
and collaboratorsi who  stressed the importance of rational cycles within the second cohomology 
of the del Pezzo complex surfaces. For instance in the case of $CP^1\times CP^1$ the middle 
cohomology is quite boringly equal to $Z+Z$, yet one axis of this lattice is selected by 
the complex geometry to 
be the root lattice of the above mentioned $A_1$ and the correspondence between  spheres
on the del Pezzo surface and D-branes on the string side \cite{V} suggested to us that 
one should combine the Weyl cone of $A_1$ and the Mori cone of the cohomology into a 
Borcherds cone associated to the simple (positive)  roots of a 
generalized Cartan matrix obtained from that of $A_2$ by replacing one of the diagonal 
elements (2) by a zero! The correspondence is best understood in this case but
more generally it is still useful \cite{JHLP1}. The intersection form on the surface is 
in this case the metric on the Cartan subalgebra of a Borcherds algebra. 

\subsection{Truncated Borcherds algebras and V-duality}
On the string/supergravity side we have  known for a while \cite{CJLP2} that there is a natural 
generalization of the Borel subgroup of U-duality  (isomorphic to the corresponding 
non-compact symmetric space and target of the scalar fields) to a solvable group 
encompassing all the p-forms and encoding
their non linear couplings but not the graviton field yet. The question was to give 
a name to this solvable group despite the absence of any reasonable classification 
of non semi-simple 
Lie algebras. It generalizes the encoding of nonlinear sigma model fields' couplings
within the structure constants of a group, to that of higher forms' couplings in the 
(super)group structure of this solvable algebra. A (p+1)-form will have degree
(p+1)  and the $Z$-graded solvable  superalgebra reduces in degree zero precisely the 
U-duality algebra or if one prefers its Borel subalgebra. There is a remarkable 
correspondence between the del Pezzo data and the string/M-theory data \cite{JHLP1}.
Two steps are left to ascend: firstly one should include gravity which only trickles 
down into this 
formalism after dimensional reduction, and secondly one must incorporate the fermions 
(this will require the enlargment of the Borel algebras to full V-duality symmetry
groups in order to allow for their ``maximal compact subgroups'' whatever this 
means to act on the fermions, but we have lots of experience even in the infinite 
dimensional case of spacetime dimension 2). 

\section{$\Lambda$-Instantons}

\subsection{Gravitational instantons}
Let us consider now a 4 dimensional Riemannian manifold and its Riemann curvature 
4-tensor $R$. It is well known \cite{EGH} that one may impose (Hodge) self-duality
on the first (or second) pair of indices, this defines the usual gravitational 
instantons which are necessarily Ricci flat and provide a nice subset of solutions 
of the second order Einstein equations. One may also require to have double 
self-duality exactly as in (\ref{SD})
\begin{equation}
R=S*R\label{ESD}
\end{equation}
where S is the dualization on the first pair of indices if * is the dualization on 
the second pair. This is equivalent to the Einstein space condition (with 
unspecified cosmological constant). There is the conformal self-duality equation 
too that guarantees the existence of a twistor space see for instance \cite{W}.

\subsection{$\Lambda$-instantons}
It seems to have gone unnoticed that there is yet another equation for any given 
value $\Lambda$ of the cosmological constant that provides what we call 
$\Lambda$-instantons \cite{JLR}. It is obtained by adding in the ordinary 
gravitational instanton equation to the Riemann curvature tensor the combination
\begin{equation}
-\Lambda/3(g_{\mu\rho}g_{\nu\sigma}-g_{\nu\rho}g_{\mu\sigma})
\end{equation}
the resulting tensor $Z_{\mu\nu\rho\sigma}$ turns out to be equal to the 
MacDowell Mansouri tensor associated to a de Sitter bundle \cite{MM}. 
The $\Lambda$-instanton equation reads simply
\begin{equation}
Z=*Z. \label{LI}
\end{equation}
It  implies the Einstein equation for that particular value of the cosmological 
constant but it is not equivqlent to it.
 	
\section{Duality in the gravitational sector}

\subsection{Near flat space}
In a nice paper \cite{HT} the dual form of 4d linearized Einstein gravity
was found to be again of the same type. The authors introduced 2 prepotentials 
and their associated pregauge invariances beyond diffeomorphism symmetry and 
showed they were interchangeable by a continuous duality rotation on shell. Even 
off shell the non-covariant  action is invariant under duality exactly 
as in the Maxwell case. Such a duality exists at the nonlinear level in the 
presence of one Killing vector field it is precisely the Ehlers symmetry, 
whereas such an isometry is not assumed 
anymore here. The prepotentials are defined by solving the hamiltonian and momentum 
constraints. 

\subsection{Near de Sitter space}
It maybe encouraging to go beyond this linear truncation to linearize around a different 
background and to try and see whether such a duality symmetry persists. Around de Sitter 
space (but the sign of the cosmological constant is not really important for local 
questions) indeed the duality rotation exchanges the relevant components of the modified 
curvature tensor $Z$, the electric part is $Z_{0m0n}$ and the magnetic part 
$1/2 \, Z_{0m}^{pq}\epsilon^{pqn}$. When the cosmological constant tends to zero the near
flat space result is recovered smoothly. 

\section{Conclusion}
We must now go nonlinear and it seems natural to expect from M-theory considerations
that the dual theory does exist and that it is worth our efforts. More specifically the 
dual diffeomorphism invariance is suggestive of a doubling of spacetime, allowing for 
some  self-duality condition that reduces the effective dimension to 4. This doubling 
is very familiar in string theory. We had no time to 
review quantum effects like quantum anomaly or NUT charge quantization.

\end{document}